# Self-blinking Dyes unlock High-order and Multi-plane Super-resolution Optical Fluctuation Imaging

Kristin S. Grußmayer[1,*], Tomas Lukes[1], Theo Lasser[2,3], Aleksandra Radenovic[1,*]

Affiliations

[1]École Polytechnique Fédérale de Lausanne, Laboratory of Nanoscale Biology, 1015 Lausanne, Switzerland

[2]École Polytechnique Fédérale de Lausanne, Laboratoire d'Optique Biomédicale, 1015 Lausanne, Switzerland

[3]Max-Planck Institute for Polymer Research, Ackermannweg 10, 55128 Mainz, Germany

*Corresponding Authors:

Kristin S. Grußmayer, email : kristin.grussmayer@epfl.ch,

Aleksandra Radenovic, email : aleksandra.radenovic@epfl.ch

# Abstract

Diffraction unlimited super-resolution imaging critically depends on the switching of fluorophores between at least two states, often induced using intense laser light and special buffers. The high illumination power or UV light required for appropriate blinking kinetics is currently hindering live-cell experiments. Recently, so-called self-blinking dyes that switch spontaneously between an open, fluorescent "on"-state and a closed colorless "off"-state were introduced. Here we exploit the synergy between super-resolution optical fluctuation imaging (SOFI) and spontaneously switching fluorophores for 2D functional and for volumetric imaging. SOFI tolerates high labeling densities, on-time ratios, and low signal-to-noise by analyzing higher-order statistics of a few hundred to thousand frames of stochastically blinking fluorophores. We demonstrate 2D imaging of fixed cells with a uniform resolution up to 50-60 nm in $6^{th}$ order SOFI and characterize changing experimental conditions. We extend multiplane cross-correlation analysis to $4^{th}$ order using biplane and 8-plane volumetric imaging achieving up to 29 (virtual) planes. The low laser excitation intensities needed for self-blinking SOFI are ideal for live-cell imaging. We show proof-of-principal time-resolved imaging by observing slow membrane movements in cells. Self-blinking SOFI  provides a route for easy-to-use 2D and 3D high-resolution functional imaging that is robust against artefacts and suitable for live-cell imaging.

Fluorescence super-resolution microscopy enables visualization of sub-diffraction structures and has led to new biological insight[1,2]. The key to overcoming the diffraction limit lies in distinguishing molecules within a diffraction-limited volume. In the majority of super-resolution imaging the switching of fluorescent emitters between two states, typically the 'on'- and "off"-state, is exploited[3,4]. The attainable resolution critically depends on the labeling and on the performance of fluorophores. Single-molecule localization microscopy (SMLM) and super-resolution optical fluctuation imaging (SOFI) rely on the stochastic and independent blinking of emitters. Light-induced (reversible) switching of organic dyes and fluorescent proteins using UV or high-intensity visible light, often in combination with special buffer systems, is commonly used in experiments[3]. Phototoxicity is a huge challenge for live cell super resolution imaging[5,6], with intensities lower than typically used in experiments leading to changes e.g. in cell proliferation. Photo-induced damage is dramatically increased at lower irradiation wavelength, which is why even low UV intensities can be detrimental[5]. A lot of efforts have been devoted to reducing phototoxicity while preserving the above-mentioned fluorophore properties. Lately, primed photoconversion of fluorescent proteins using blue and infrared light was used for SMLM[7,8].

Recently, self-blinking dyes based on an intramolecular spirocyclization reaction of rhodamine derivatives compatible with live-cell imaging were introduced[9]. Spontaneous fluorescence fluctuations occur due to the chemical structure, and the equilibrium between open (fluorescent "on"-state) and closed (colorless "off"-state) form can be influenced by the environment[9–11](see schematic in Figure S1). The intrinsic blinking kinetics can be tuned by changing the nucleophilicity of the intramolecular nucleophile and the electrophilicity of the fluorophore scaffold. Screening of many compounds initially revealed HM-SiR as the only candidate that achieved the sparse blinking necessary for SMLM. High emitter densities in combination with inappropriate switching rates can easily lead to imaging artifacts[12] in localization based super-resolution microscopy. Super-resolution optical fluctuation imaging[13] performs higher-order statistical analysis, i.e. spatio-temporal cross-cumulants, of the time-series of fluctuating fluorophores instead of localizing individual emitters. Resolution increases with increasing cumulant order of the analysis further referred to as "SOFI order". For example, 4$^{th}$ order SOFI achieves up to 4 times resolution enhancement of the input widefield image. SOFI tolerates higher labeling densities, on-time ratios and lower signal-to-noise and uses only a few hundred-thousand frames for cumulant calculation[14,15]. In addition to resolution enhancement, combining 3 orders of SOFI analysis can be used for molecular parameter estimation of the on-time ratio, i.e. the probability of the fluorophore to reside in the on-state, the molecular brightness and density [16]. The combination of SOFI together with self-blinking dyes is particularly attractive for high-order, high-resolution live cell imaging.

Here, we exploit the synergy between SOFI and the commercially available dye Abberior FLIP 565. Abberior FLIP 565 is a reversibly switching spiroamide compound[17,18]. Our experiments achieve high-order, up to 50-60 nm high-resolution SOFI functional 2D imaging and volumetric SOFI with up to 29 planes in 4$^{th}$ order cumulant analysis. We show that the spatially homogeneous fluorophore blinking kinetics lead to uniform resolution increase in the field-of-view. Using low laser excitation intensities, we demonstrate the potential of self-blinking SOFI for live cell imaging.

## Results

*Functional 2D Imaging*

First, we investigated the performance of Abberior FLIP 565 for high-order super-resolution optical fluctuation imaging in fixed cells using standard widefield fluorescence illumination. Microtubules in COS-7 cells were immunostained at high fluorophore densities and subsequently imaged under different excitation intensities and buffer conditions. Here, we focus on the self-blinking properties of Abberior FLIP 565 molecules via thermal relaxation in densely labeled samples. Figure 1a and b shows exemplary SOFI analysis of the microtubule cytoskeleton of a cell that is spread across the field of view. Efficient background suppression and an increase in optical sectioning for higher-order SOFI analysis are apparent (see Figure S2 for full field of view). We estimated the resolution of SOFI images using image decorrelation analysis[19] and verified that it increases with the cumulant order. We achieve excellent resolution down to about 50-60 nm at 6$^{th}$ order analysis using low illumination intensities <0.17 kWcm$^{-2}$ 561nm. The self-blinking of the dye results in homogeneous switching kinetics, independent of the Gaussian illumination profile of the microscope. The on-time ratio map ($\rho = \frac{\tau_{on}}{\tau_{on}+\tau_{off}}$ with the characteristic lifetime $\tau_{on/off}$ of the "on/off"-state of the fluorophore, see Figure 1c) calculated from SOFI data confirms this expected behavior, leading to uniform spatial resolution across the image (see Figure S3). Under the tested illumination conditions, we can collect thousands of frames with negligible apparent photobleaching (for the examples see Figure 1d). A dedicated increase in excitation intensities leads to a slight increase in photobleaching rate and is reflected accordingly in the SOFI brightness readouts (see Figure 1e)). We also show that the on-time ratio increases about two-fold when PBS buffer is exchanged for 50% glycerol (see Figure 1f) and Figure S1). This shift of the equilibrium towards the open, fluorescent form is in line with previous observations. Similar dyes have a lifetime $\tau_{on}$ of the "on"-state of milliseconds in polar solvents which increases up to hours in polyvinyl alcohol[18]. The observed on-time ratio in the low percent regime, together with high density labeling already poses a significant challenge resulting in overlapping emitters for traditional SMLM, but is perfectly compatible with high quality SOFI. Next, we explored the feasibility of high-order self-blinking 3D SOFI based on multi-plane imaging.

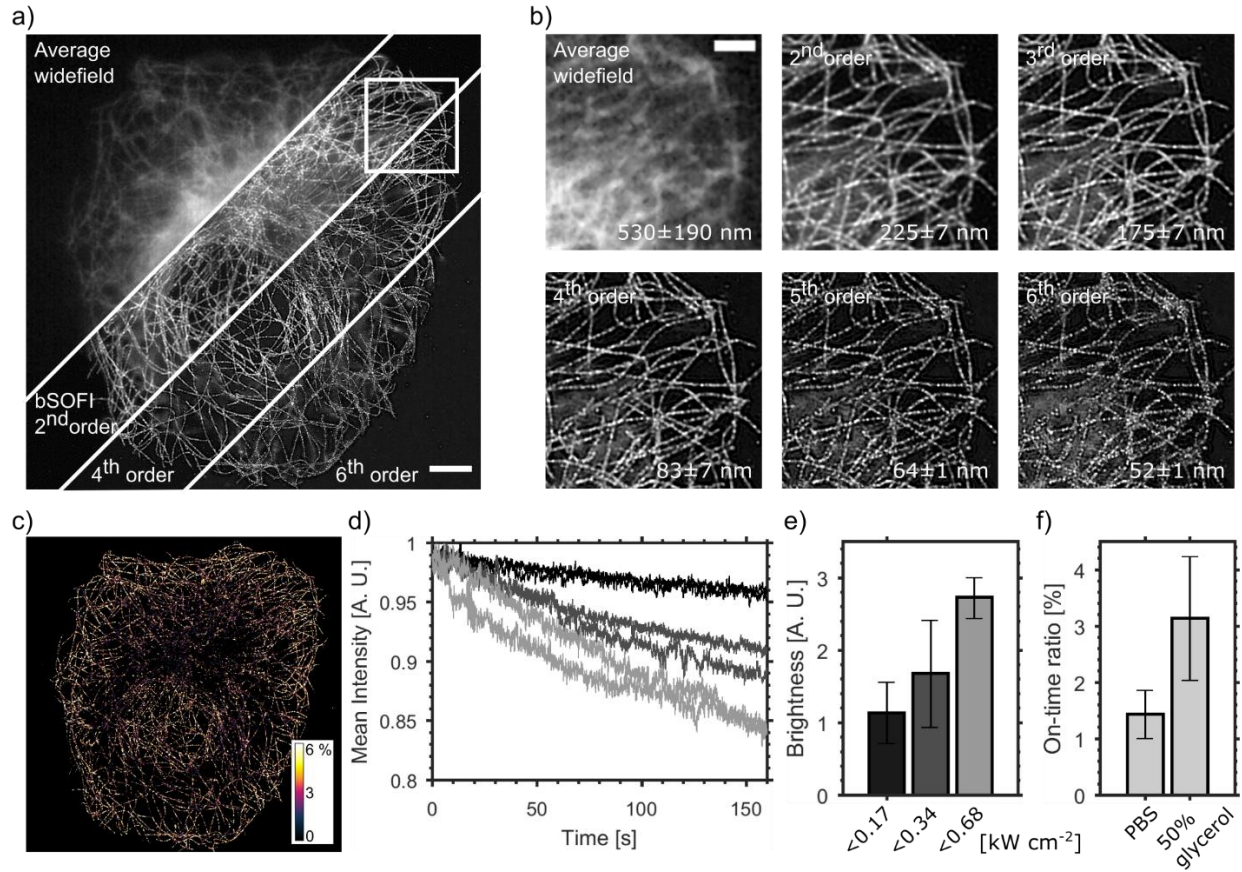

Figure 1 *High-order 2D self-blinking SOFI and functional imaging.* Resolution increases with increasing cumulant order. COS-7 cells immunostained for microtubules with Abberior Flip 565. a) Widefield (average of the image sequence), $2^{nd}$, $4^{th}$ and $6^{th}$ order SOFI. 50% glycerol in PBS with <0.17 kW cm$^{-2}$ 561nm excitation, 8000 frames at 20 ms exposure time. Scale bar 5µm b-) $2^{nd}$- $6^{th}$ order SOFI, close-up of areas marked in a). Resolution estimate by decorrelation analysis for three different whole cells (average ± standard deviation, see Materials & Methods). c) on-time ratio map $\rho = \frac{\tau_{on}}{\tau_{on}+\tau_{off}}$ of the cell in a). d) Photobleaching under different laser intensities, <0.17/0.34/0.68 kW cm$^{-2}$ 561nm excitation (black/dark grey/light grey). Comparison of brightness e) and on-time ratio f) under different imaging conditions.

*Multi-plane Imaging*

Having established imaging conditions in 2D, we next aimed for volumetric self-blinking SOFI. We previously established 3D SOFI multi-plane imaging using a dedicated image-splitting prism in the detection path[20]. The simultaneous volumetric image acquisition allows us to exploit cross-cumulant analysis in all spatial dimensions. This leads to increased resolution and sampling also along the z-dimension, thereby creating virtual planes in addition to the physically acquired ones.

Here, we introduce biplane 3D SOFI imaging that does not require customized parts and poses minimal requirements on optical design of the microscope. Biplane imaging was implemented as an extension of the 2D system described above by introducing a 50:50 beam splitter and a second synchronized camera (see inset Figure 2a and Supplementary Figure S4). The position of the additional camera with respect to the tube lens was offset compared to the other camera such that the corresponding focal planes were 800nm apart in object space. This inter-plane distance allows appropriate sampling of the axial point spread function. The emission light is equally split between the cameras, leading to half the photon budget for SOFI cumulant analysis for given excitation intensities compared to the 2D case. We extended our 3D SOFI image processing to $4^{th}$ order cumulant calculation and adapted it to biplane imaging using a workflow similar to what we previously described[20]. We determined the interplane distance by scanning fluorescent beads through the focus. For ease of comparison, we image the same structures to showcase the different SOFI modalities. Self-blinking dyes easily allow the computation of $4^{th}$ order SOFI at low excitation intensities, showing how microtubules trace the 3D shape and outline of the cell in Figure 2a and b. Again, background reduction, sectioning and resolution improvement can be qualitatively seen in the z-projections (see Figure S5 for an exemplary 2D plane).

In the next step, we tested larger volume acquisitions of microtubules enclosing the nucleus using the existing 8-plane prism-based setup with 350nm interplane distance (see inset Figure 2c, Supplementary

Figure S4 and [20]). Fluorescence fluctuations of Abberior Flip 565 again allow computation of SOFI up to 4th order, resulting in 3D images containing 29 planes. SOFI image improvement can be qualitatively seen in the z-projections; an exemplary 2D plane is shown in Figure S6. Since the signal is divided among 8 planes and the excitation efficiency of the dye with the laser on this setup was only 44%, we had to work at higher illumination intensities <3.5 kWcm$^{-2}$ 532nm as compared to 2D and biplane imaging. These measurements are the first successful demonstrations of 4th order multi-plane SOFI. Obtaining three SOFI cumulant orders is required for calculating molecular parameter estimates such as on-time ratio and molecular densities. Our measurements thus put into perspective quantitative functional imaging in 3D.

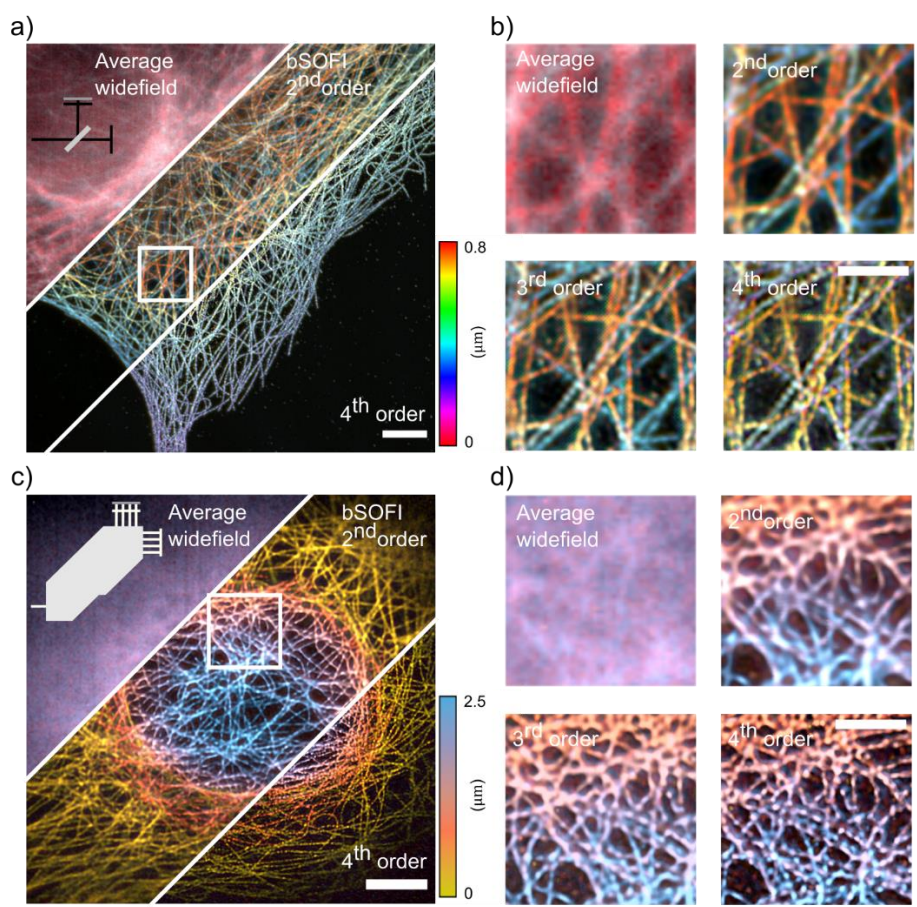

*Figure 2 High-order multi-plane 3D self-blinking SOFI.* COS-7 cells immunostained for microtubules with Abberior Flip 565. a) Biplane imaging: Widefield (average of the image sequence), 2nd & 4th order SOFI. 50% glycerol in PBS with <0.68 kW cm$^{-2}$ 561nm excitation, 6000 frames at 50 ms exposure time. Scale bar 5µm. b) close-up of areas marked in a). Scale bar 2µm. c) 8-plane imaging: Widefield (average of the image sequence), 2nd & 4th order SOFI, 50% glycerol in PBS with <3.5 kW cm$^{-2}$ 532 nm excitation, 8000 frames at 50 ms exposure time. Scale bar 5µm. d) close-up of areas marked in c). Scale bar 2µm.

*Live-cell Imaging*

SOFI using Abberior Flip 565 self-blinking requires only low illumination intensities <0.17 kW cm$^{-2}$ 561nm (see 2D and biplane imaging results above). This minimizes light-induced changes in cellular behavior (below threshold determined for 514nm light in [5]) and limits photobleaching, making it compatible with longer term observation in living cells. We labeled U2OS cells using the lectin wheat germ agglutinin conjugated to Abberior Flip 565, targeting N-acetyl-D-glucosamine and sialic acid residues that are common in membrane glycoproteins (see Figure S7 for fixed COS-7 cells with wheat germ agglutinin labeling). Figure 4a shows the average of a 400s sequence of a 2D slice slightly above the coverslip, highlighting the striking sectioning capability of SOFI. The leading edge of the cell, presumably showing membrane protrusions or ruffling, as well as staining of intracellular membranes, can be seen. Time-resolved 4$^{th}$ order SOFI analysis (see Figure 4b and Figure S8) reveals slow dynamics of membrane rearrangements. There was no need for rescaling of the colormap indicating that photobleaching was negligible.

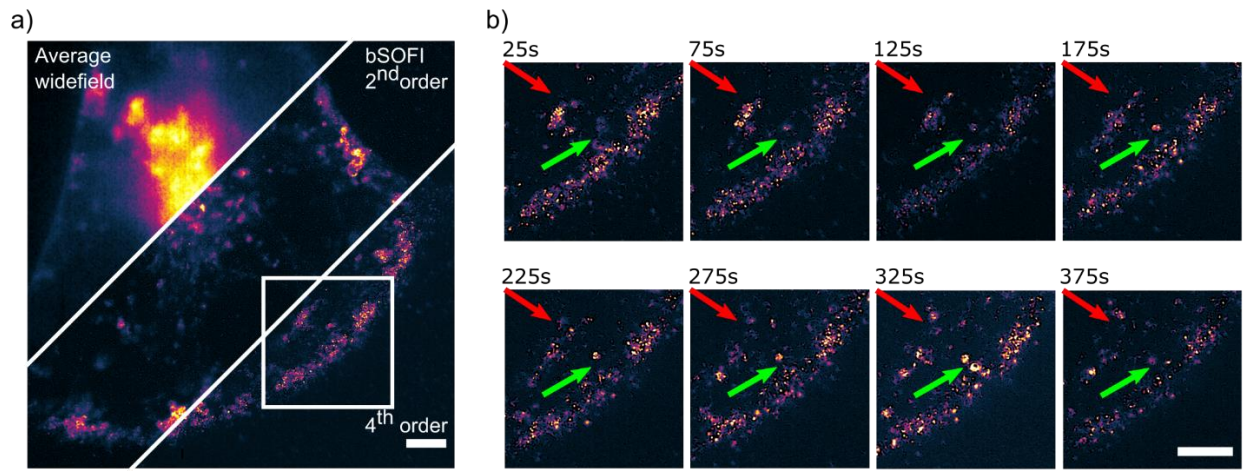

*Figure 3 Slow dynamics of wheat germ agglutinin-Abberior Flip 565 stained U2OS cell.* a) Widefield (average of the image sequence), 2$^{nd}$, and 4$^{th}$ order SOFI of the whole imaging sequence (8000 frames). 2D imaging in HBSS with <0.17 kW cm$^{-2}$ 561nm excitation. b) 4$^{th}$ order SOFI close-up of areas marked in a) at different time points with a sliding window of 50 s (1000 frames), every second time point is shown (see Figure S7 for all data). Scale bar 5μm.

For future investigations of cell biology, careful investigation of the time-resolution necessary to reveal the respective dynamic processes under investigation are mandatory. In general, less frames can be used for second order analysis, sacrificing the possibility of functional imaging via estimation of molecular parameters (at least three SOFI orders are necessary[16]) to increase the time-resolution. SOFI achieves super-resolution already when analyzing less frames than needed for localization microscopy, potentially offering higher time-resolution[14,15]. We did not pursue maximizing the imaging speed in this application. Here, our aim was to use the commercially available Abberior Flip dye and show high-order live cell SOFI using organic dyes.

## Summary and Discussion

We demonstrated super-resolution imaging in 2D and 3D in fixed and live cells, exploiting the combined advantages of self-blinking dyes and cumulant analysis. Our results show that the blinking behavior of Abberior STAR 565 allows high-order super-resolution optical fluctuation imaging at up to 50—60 nm resolution with organic dyes at extremely low illumination intensities <0.17 kW cm$^{-2}$ 561nm. In addition, we investigated the photophysical behavior of the dye under changing illumination and buffer conditions using SOFI based brightness and on-time ratio estimation in 2D. Our analysis shows that self-blinking SOFI delivers uniform resolution across the field-of-view despite inhomogeneous illumination, facilitated by the alternative switching mechanism that is insensitive to spatial variation in the illumination intensity.

Biplane data acquisition results only in a limited reduction in signal brightness and allows imaging of a thin section of cells. This first demonstration of biplane cross-cumulant analysis highlights easy-to-implement 3D imaging with low setup complexity. It is ideal for investigating e.g. proteins in the basal membrane of cells, since it is known that there are limitations in quantitative imaging due to 2D projections induced by vesicle-like structures, invaginations and overlapping membranes[12]. We extended multi-plane cross-cumulant analysis to 4$^{th}$ order and show that we can achieve high-quality, high-order volumetric imaging of 29 (virtual) planes in our 8 (real)- plane prism microscope. This provides the starting point for volumetric SOFI-based estimation of molecular parameters such as molecular densities, brightness and the ratio of dyes in the on-state. Our study thus paves the way for microenvironment-related functional super-resolution microscopy using the fluorophore properties as indicators.

Finally, the low laser excitation intensities needed for self-blinking SOFI and the robustness against high molecule densities are ideal for live cell imaging. We show proof-of-principal high-order imaging by observing slow movements in U2OS cells. Probabilities for the fluorphores to be in the on-state on the order of one per-cent (typical for spontaneously blinking dyes[8,9,16,21]), especially paired with high labeling densities were previously shown to pose serious challenges for single-molecule localization microscopy[12,22]. A higher on-time ratio and fast blinking speed are ideal for SOFI, thus a wide range of already synthesized self-blinking dye compounds[9,23] not suitable for SMLM are good candidates for imaging.

Self-blinking SOFI provides a route for easy-to-use 2D and 3D high-resolution functional imaging that is robust against artefacts and perfectly compatible with live-cell imaging.

# Materials and Methods

*Microscope for 2D and Biplane Imaging*

Data was acquired on a home-built widefield microscope as described in [19]. Briefly, four illumination laser lines, i.e. 200 mW 405 nm laser (MLL-III-405-200mW), a 1 W 635 nm laser (SD-635-HS-1W, both Roithner Lasertechnik), a 350 mW 561 nm laser (Gem561, Laser Quantum) and a 200mW 488nm laser (iBEAM-SMART-488-S-HP, Toptica Photonics) are collimated, expanded and combined with dichroic filters and focused in the back focal plane of the objective (Nikon SR Plan Apo IR 60× 1.27 NA WI). The fluorescence light is filtered using a combination of a dichroic mirror and a multiband emission filter (Quad Line Beamsplitter R405/488/561/635 flat and Quad Line Laser Rejectionband ZET405/488/561/640, both AHF Analysetechnik) and subsequently focused on two synchronized sCMOS cameras (ORCA Flash 4.0, Hamamatsu; back projected pixel size of 108 nm). The sample position is controlled in X and Y by a Scan-plus IM 120x80 (Marzheuser) and in Z by a Nano-Z piezo nanopositioner (Mad City Labs). All acquisitions were performed using Micromanager. For 2D imaging using the "transmission" camera, the beamsplitter in the detection path was removed and for imaging using the "reflection" camera the dichroic beamsplitter HC BS 640 imaging (AHF Analysentechnik) was in place. For biplane imaging, a 50:50 beamsplitter (BSW26R, Thorlabs) was used.

*Microscope for Multi-plane Imaging*

Data was acquired on a home-built widefield microscope with 8-plane detection described in detail in [20]. Briefly, four illumination laser lines, a 120 mW 405 nm laser (iBeam smart, Toptica), an 800 mW 635 nm laser (MLL-III-635, Roithner Lasertechnik), and an 800 mW 532 nm laser (MLL-FN-532, Roithner Lasertechnik) are collimated, expanded and combined with dichroic filters and focused into the back focal plane of the objective (Olympus UPLSAPO 60XW 1.2 NA). The fluorescence light is filtered using a dichroic mirror and quad band emission filter (zt405/488/532/640/730rpc and ZET 405/488/532/640m Chroma).

The detection path is arranged as a sequence of four 2-f configurations to ensure image–object space telecentricity. An image splitting prism placed behind the last lens directs the light into eight images recorded by two synchronized sCMOS cameras (ORCA Flash 4.0, Hamamatsu; back projected pixel size of 111 nm). For translating the sample, the microscope is equipped with piezoLEGS stage (3-PT-60- F2,5/5) and Motion-Commander-Piezo controller (Nanos Instruments GmbH).

***Super-resolution optical fluctuation imaging data and processing***

*Table 1* Summary of samples, data acquisition and processing parameters.

|  | sample | # frames total/subsequence | Exposure time [ms] | illumination intensity [kW cm$^{-2}$] |
|---|---|---|---|---|
| Figure 1a-c | microtubules in fixed COS-7 cells, 50% glycerol in PBS | 8000/2000 | 20 | <0.17 561 nm |
| Figure 1d | See above | 8000 | 20 or 50 | <0.17/0.34/0.68 561 nm |
| Figure 1e | microtubules in fixed COS-7 cells, 50% glycerol in PBS or PBS | 8000/2000 | 20 or 50 | <0.17/0.34/0.68 561 nm |
| Figure 1f | See above | See above | 20 or 50 | 561 nm |
| Figure 2a+b | microtubules in fixed COS-7 cells, 50% glycerol in PBS | 6000/2000 | 50 | <0.68 561 nm |
| Figure 2c+d | See above | 8000/2000 | 50 | <3.5 532 nm |
| Figure 3a | WGA labeled U2OS cells in HBSS | 8000/1000 | 50 | <0.17 561 nm |
| Figure 3b | See above | 1000/1000 | 50 | See above |

*2D* data

Cumulant calculation and the subsequent determination of molecular parameters (on-time ratio, brightness) was performed as described in [15], except that the brightness response was linearized by taking the n-th root of the deconvolved cumulant image according to the original bSOFI formulation (similar to

3D image processing). 2D drift-correction was performed using cross-correlation between different SOFI subsequences before averaging.

*3D* data

The workflow for multi-plane data analysis was essentially as described in [20]. We extended the 3D cross-cumulant analysis to 4th order. For biplane data, coregistration was refined using cross-correlation between the two image planes. Due to the small z-extend of the biplane data we essentially performed 2D deconvolution.

### *Resolution estimation*

We estimate the resolution based on image decorrelation analysis as described in ref [19], using the ImageJ plugin with a setting of 20 Gaussians. A resolution worse than expected for widefield imaging is determined for the averaged image due to bad signal-to-noise and missing lateral drift-correction. The resolution map was calculated using the freely available Matlab code with the same settings as for the ImageJ plugin.

### *Cell preparation*

*Microtubule staining*

Fixed COS-7 cells were prepared as described in [19]. Primary anti-tubulin antibody (clone B-5-1-2 ascites fluid, Sigma-Aldrich) was used at 1:100-1:200 dilution, secondary donkey anti-mouse-AbberiorFlip565 (preparation see below) was used at 1:100-1:200 dilution.

COS-7 cells were cultured at 37 °C and 5 % $CO_2$ using DMEM high glucose w/o phenol red (4.5 g $l^{-1}$ glucose) supplemented with 4 mM L-gluthamine, 10 % fetal bovine serum and 1× penicillin-streptomycin (all gibco®, Thermo Fisher Scientific). Cells were seeded in Lab-tek® II chambered cover slides (nunc) or on 18 mm high-precision No. 1.5 borosilicate coverslips (Marienfeld) in 12 well plates (Thermo Fisher Scientific) 1-2 days before fixation in DMEM high glucose w/o phenol red (see above).

The fixation and labelling protocol is similar as described by Chazeau et al.[24]. Cells were washed twice in pre-warmed DMEM w/o phenol red following 90 s incubation with extraction buffer (microtubule stabilizing buffer 2 (MTSB2: 80 mM PIPES, 7 mM MgCl2, 1 mM EGTA, 150 mM NaCl, 5 mM D-glucose adjust pH to 6.8 using KOH) with freshly added 0.3 % Triton X-100 (AppliChem) and 0.25 % glutaraldehyde (stock solution 50 % electron microscopy grade, Electron Microscopy Sciences). Immediately afterwards, pre-warmed 4 % paraformaldehyde (PFA) in PBS was incubated for 10 min at room temperature (RT). Cells were then washed three times for 5 min each with 1× PBS and stored in 50 % glycerol in 1× PBS at 4 °C or the immunostaining protocol was continued. Next, a freshly prepared solution of 10mM NaBH4 in 1× PBS was incubated on the cells for 7 minutes followed by one quick wash in 1× PBS, and two washes 10 min 1× PBS on an orbital shaker. Cells were permeabilized in PBS with 0.25 % Triton X-100 for 7 min followed by blocking with blocking buffer (BB: 2 % (w/v) BSA, 10 mM glycine, 50 mM ammonium chloride $NH_4Cl$ in PBS pH 7.4 for 60 min at RT or overnight at 4 °C.

The blocked samples were incubated with primary anti-tubulin antibody (clone B-5-1-2 ascites fluid 1:100-1:200 dilution, Sigma-Aldrich) in BB for 60 min at RT. Cells were then washed three times for 5 min each with BB, followed by incubation with donkey anti-mouse-AbberiorFlip565 (preparation see below at 1:100-1:200 dilution) for 60 min at RT. This and all subsequent steps were performed in the dark. Cells were again washed three times for 5 min each with BB and incubated for 10 min post-fixation with 2 % PFA in 1× PBS followed by three 5 min washes with PBS. Cells were imaged immediately or stored in 50 % glycerol in 1× PBS at 4 °C until SOFI imaging.

*Wheat germ agglutinin staining:*

COS-7 and U2OS cells were cultured at 37 °C and 5 % $CO_2$ using DMEM high glucose w/o phenol red (4.5 g $l^{-1}$ glucose) supplemented with 4 mM L-gluthamine, 10 % fetal bovine serum and 1× penicillin-streptomycin (all gibco®, Thermo Fisher Scientific). Cells were seeded in Lab-tek® II chambered cover slides (nunc) or on 18 mm high-precision No. 1.5 borosilicate coverslips (Marienfeld) in 12 well plates (Thermo Fisher Scientific) for 1-2 days. Cells were washed twice in pre-warmed HBSS following 10 min incubation at 37 °C and 5 % $CO_2$ with 30 µg $ml^{-1}$ WGA-Abberior FLIP 565 diluted in HBSS. Immediately afterwards, cells were washed twice with HBSS and imaged on the microscope at RT for up to 30 min or pre-warmed 4 % paraformaldehyde (PFA) in PBS was incubated for 10 min at room temperature (RT). Fixed cells were then washed three times for 5 min each with 1× PBS and stored in 50 % glycerol in 1× PBS at 4 °C.

**Preparation of labeled proteins**

2 mg $ml^{-1}$ donkey anti-mouse (H+L) highly cross-adsorbed antibody (Invitrogen) and 2 mg $ml^{-1}$ wheat germ agglutinin (WGA, Vector Labs) was incubated with Abberior FLIP 565-NHS (Abberior) at a ratio of 1: 6 for 1h at RT while shaking with the pH raised to 8.3 using sodium bicarbonate. The mixture was purified using illustra NAP Columns (GE Healthcare) according to manufacturer's instructions and eluted with slightly acidic PBS to recover the labeled antibody at neutral pH. The protein concentration was estimated by absorption spectrometry to <1.5 mg $ml^{-1}$ for donkey anti-mouse Abberior FLIP 565 and <2.5 mg $ml^{-1}$ for WGA- Abberior FLIP 565.

## Acknowledgements

K. S. G. acknowledges the support from the Horizon 2020 Framework Program of the European Union under the Marie Skłodowska-Curie Grant Agreement No. [750528] and thanks the NVIDIA Corporation for the donation of a Titan Xp GPU. A.R. and T.La. acknowledge support from the Horizon 2020 Framework Programme of the European Union ADgut [Grant No. 686271].

## Author Contributions

K.S.G. designed the research, performed and analyzed the experiments and contributed to code adaptation. T.L. extended and adapted 3D SOFI image processing. K.S.G. wrote the manuscript with input from all authors. A.R. supervised the research.

## Competing interests

The author(s) declare no competing interests.

## Data Availability

The datasets generated during and/or analyzed during the current study are available from the corresponding author(s) on reasonable request.

**Graphic for Table of Contents only**

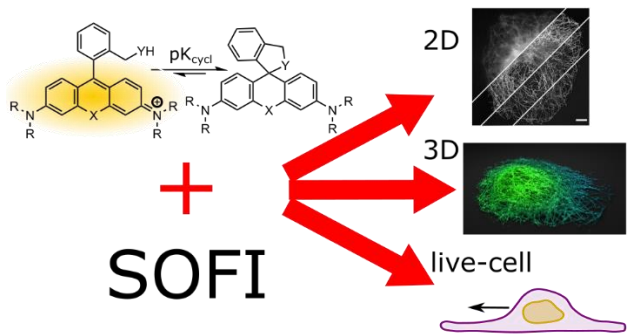

# Supporting Information: Self-blinking Dyes unlock High-order and Multi-plane Super-resolution Optical Fluctuation Imaging

Kristin S. Grußmayer[1,*], Tomas Lukes[1], Theo Lasser[2,3], Aleksandra Radenovic[1,*]

Affiliations

[1]École Polytechnique Fédérale de Lausanne, Laboratory of Nanoscale Biology, 1015 Lausanne, Switzerland

[2]École Polytechnique Fédérale de Lausanne, Laboratoire d'Optique Biomédicale, 1015 Lausanne, Switzerland

[3]Max-Planck Institute for Polymer Research, Ackermannweg 10, 55128 Mainz, Germany

*Corresponding Authors:

Kristin S. Grußmayer, email : kristin.grussmayer@epfl.ch,

Aleksandra Radenovic, email : aleksandra.radenovic@epfl.ch

## Microtubule imaging in 2D

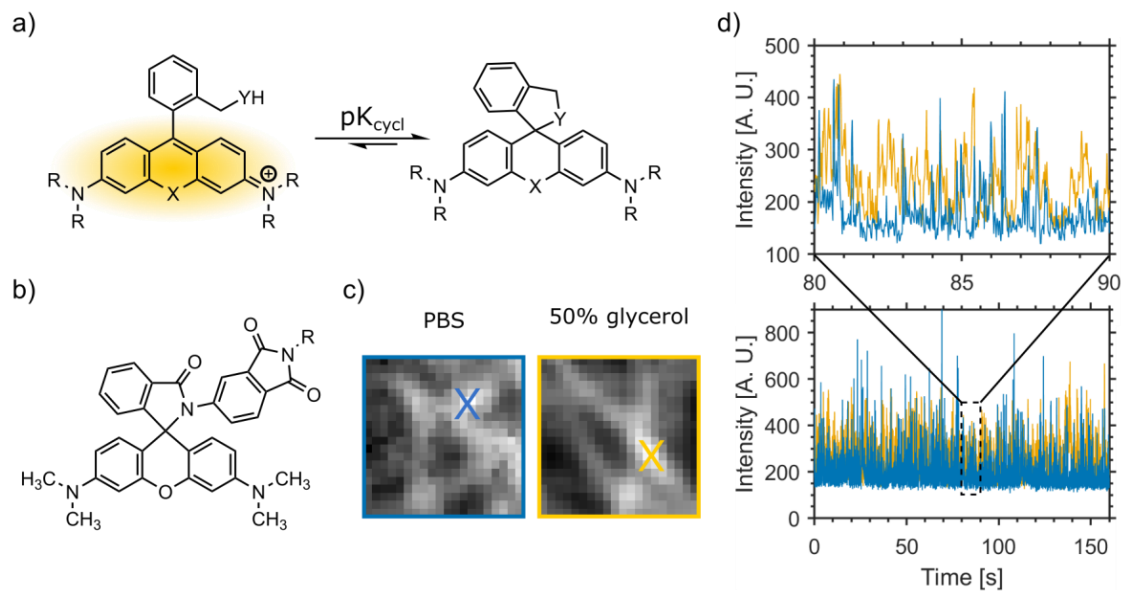

*Figure 1 Spontaneous switching of self-blinking dyes.* a) Intramolecular spirocyclization reaction of rhodamine based dyes with equilibrium constant $pK_{cycl}$. The equilibrium between the open, fluorescent "on"-state and the closed colorless "off"-state can be influenced by the fluorophore and the intramolecular nucleophiles Y[1]. b) Non-fluorescent lactone form of Abberior Flip 565 (from Abberior and [2]), c) Average widefield image of COS-7 cells immunostained for microtubules with Abberior Flip 565 imaged in PBS or 50% glycerol in PBS, representative regions (2.16 µm x 2.16 µm) with similar microtubule crossing in the same sample. Data contained in Figure 1d-f. <0.34 kW cm$^{-2}$ 561nm excitation, 8000 frames at 20 ms exposure time, d) Intensity time trace of the pixels indicated in c)(blue PBS, yellow 50% glycerol in PBS).

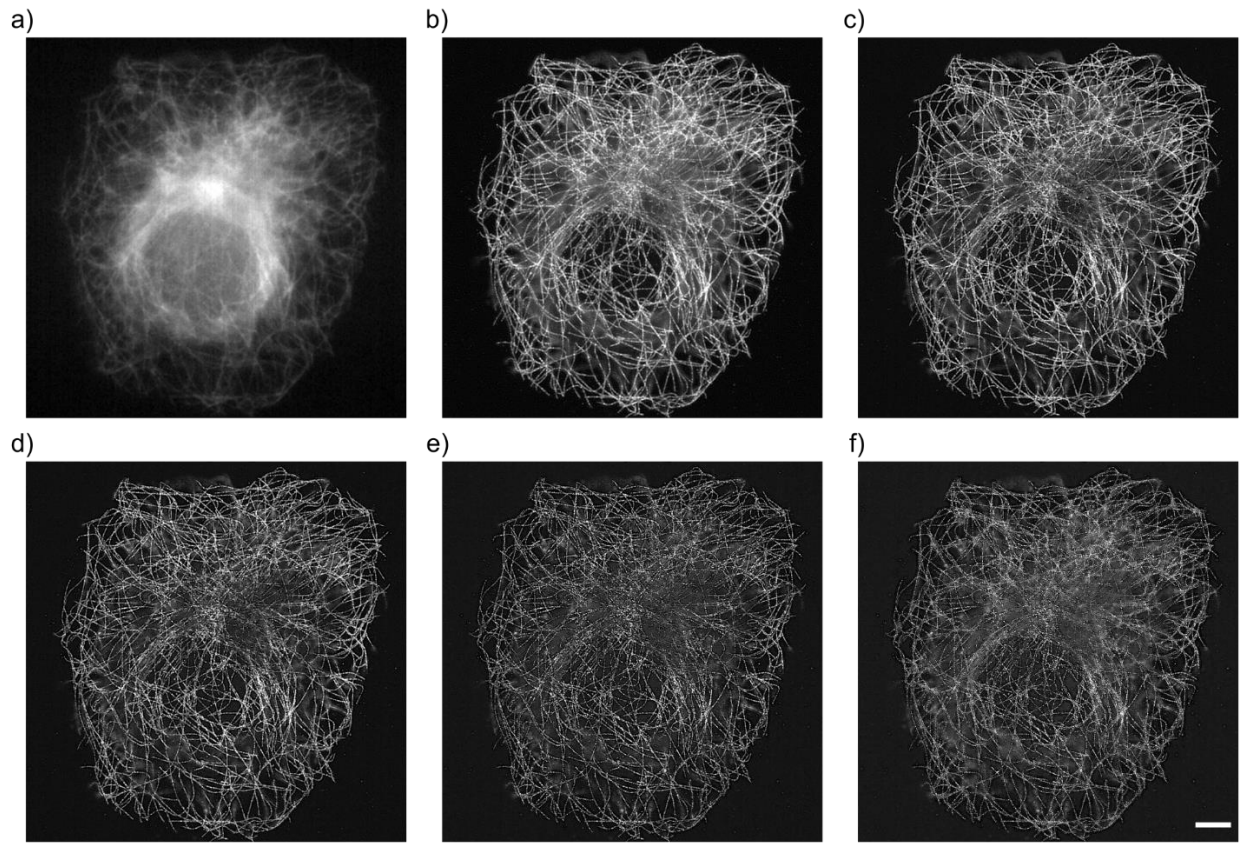

*Figure S2 High-order 2D imaging of microtubules labeled with Abberior FLIP 565 in COS-7 cells.* a-f) Pseudo widefield (average of the image sequence), 2nd, 3rd, 4th, 5th and 6th order SOFI. 50% glycerol in PBS with <0.17 kW cm$^{-2}$ 561nm excitation, 8000 frames at 20 ms exposure time. Data from Figure 1. Scale bar 5µm.

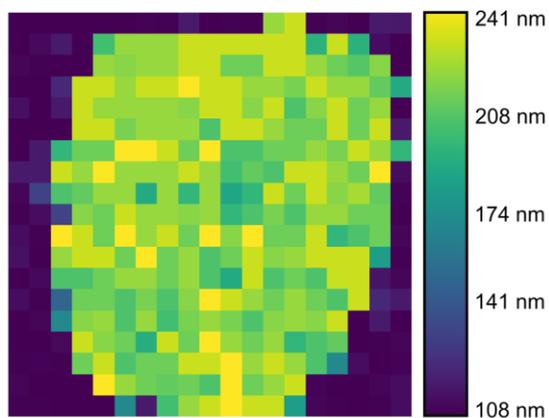

*Figure S3 Resolution map of microtubules labeled with Abberior FLIP 565 in a COS-7 cell.* Second order SOFI image processed using image decorrelation analysis to estimate the spatial frequency content with a tile size of 2.7µm x 2.7µm (50 pixels x 50 pixels). The resolution estimates on the order of the projected pixel size (108nm) in the area without signal outside of the cell are meaningless. Data from Figure 1.

## Microtubule imaging in 3D

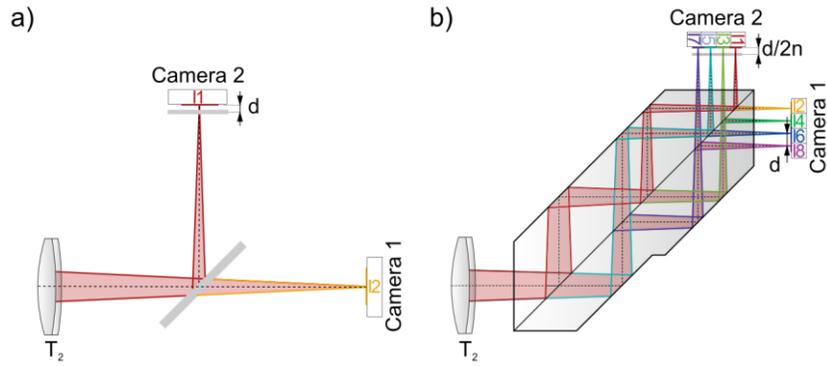

*Figure S4 Multiplane detection.* a) Biplane detection using a 50:50 beamsplitter, b) 8-plane detection using a custom image splitting prism with a 50:50 beamsplitting coating (for full description of the setup see[3]). Colors indicate the detection channels that are conjugated to object-planes at different z-positions.

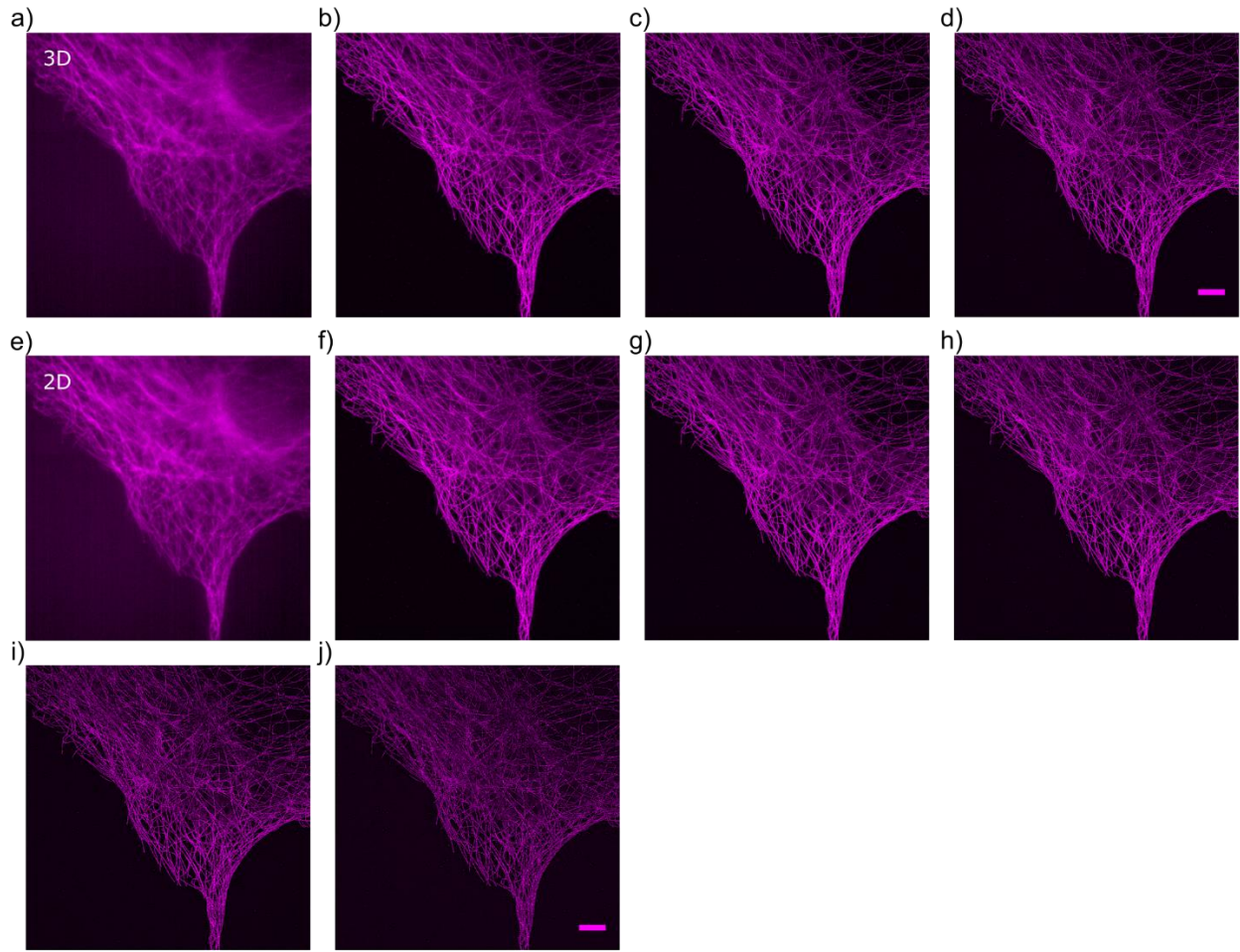

*Figure S5 Exemplary plane of high-order biplane 3D imaging.* COS-7 cells immunostained for microtubules with AbberiorFlip 565 same data as Figure 2. a-d) Widefield (average of the image sequence), $2^n$, $3^{rd}$ & $4^{th}$ order SOFI bottom plane of Figure 2a. e-j) Widefield (average of the image sequence), $2^n$, $3^{rd}$, $4^{th}$, $5^{th}$ & $6^{th}$ order SOFI processed using the 2D code that extends to higher orders. 50% glycerol in PBS with <0.68 kW cm$^{-2}$ 561nm excitation, 6000 frames at 50 ms exposure time. Scale bar 5 µm.

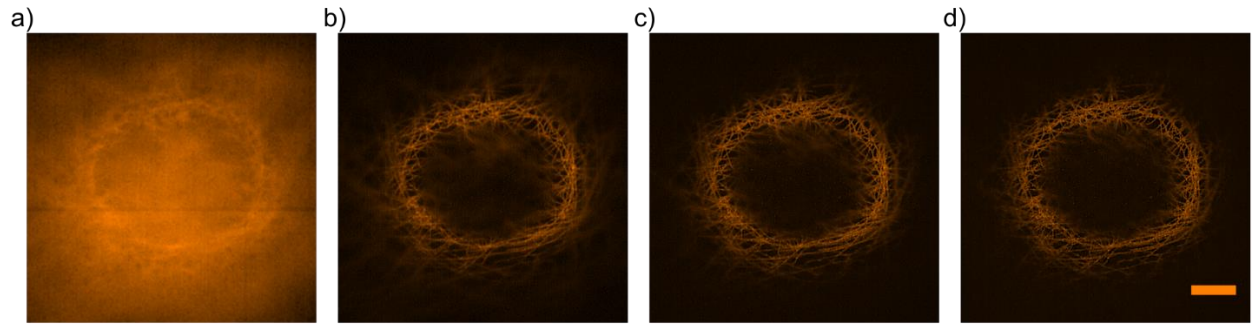

*Figure S6 Exemplary plane of high-order 8-plane 3D imaging.* COS-7 cells immunostained for microtubules with Abberior Flip 565 same data as Figure 2. a-d) Widefield (average of the image sequence), $2^n$, $3^{rd}$ & $4^{th}$ order SOFI. 50% glycerol in PBS with <3.6 kW cm$^{-2}$ 532nm excitation, 8000 frames at 50 ms exposure time. Scale bar 5 µm.

## *Wheat germ agglutinin imaging*

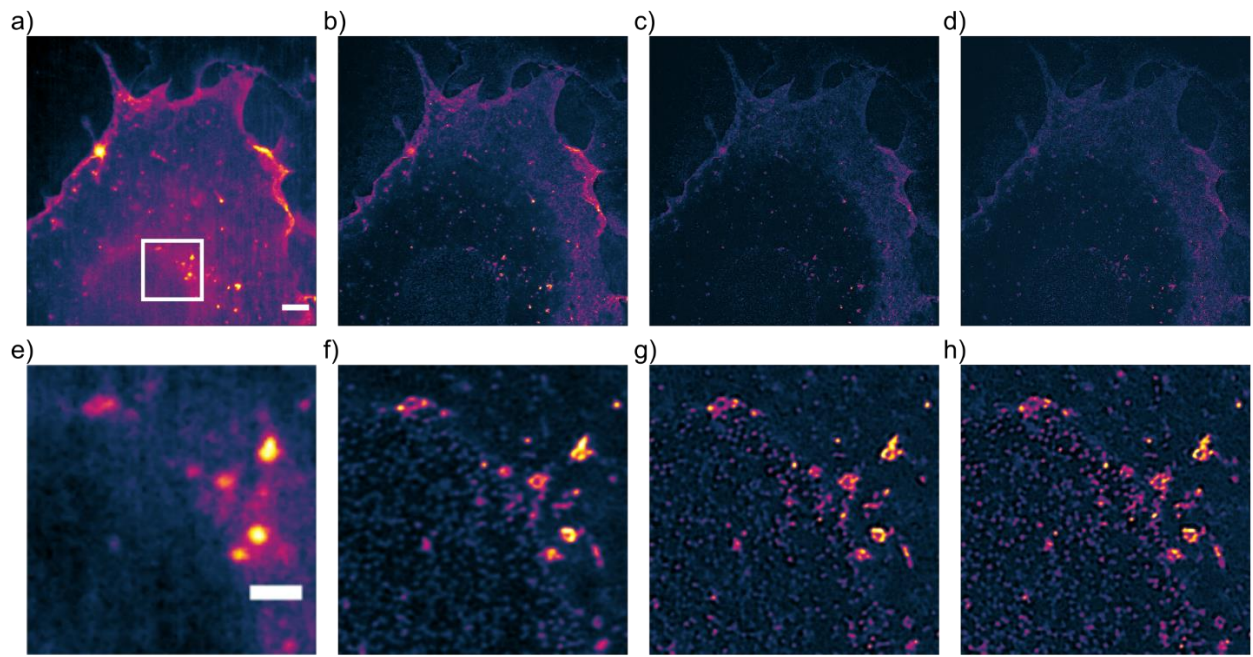

*Figure S7 High-order 2D imaging of COS-7 cells stained with wheat germ agglutinin-Abberior FLIP 565 in fixed cells.* a-d) Widefield (average of the image sequence), $2^{nd}$, $3^{rd}$ & $4^{th}$ order SOFI. Scale bar 5µm. e-h) close-up of areas marked in a), the central subunit of nuclear pore complexes can be seen as puncta in the lower nuclear membrane. Scale bar 2µm. 50% glycerol in PBS with <17 kW cm$^{-2}$ 561 nm excitation, 8000 frames at 50 ms exposure time.

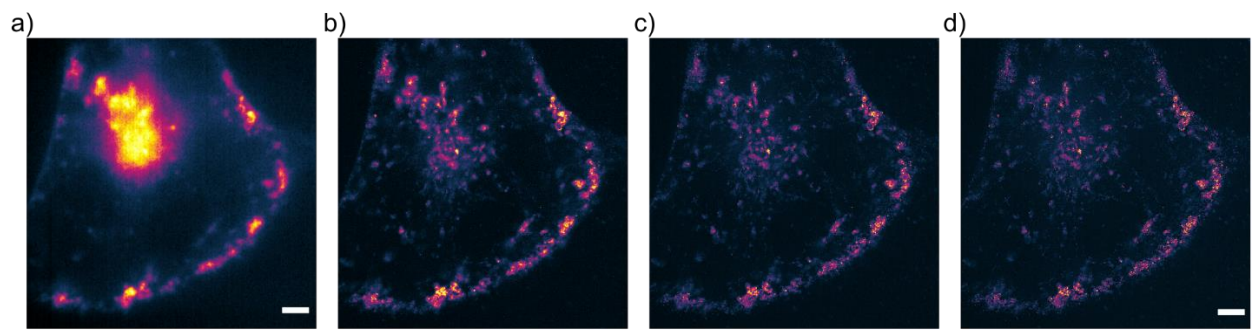

*Figure S8 High-order 2D live imaging of wheat germ agglutinin-Abberior FLIP 565 stained U2OS cell.* a-d) Widefield (average of the image sequence), 2nd, 3rd and 4th order SOFI of the whole imaging sequence (8000 frames). 2D imaging in HBSS with <0.17 kW cm$^{-2}$ 561nm excitation. Scale bar 5μm.

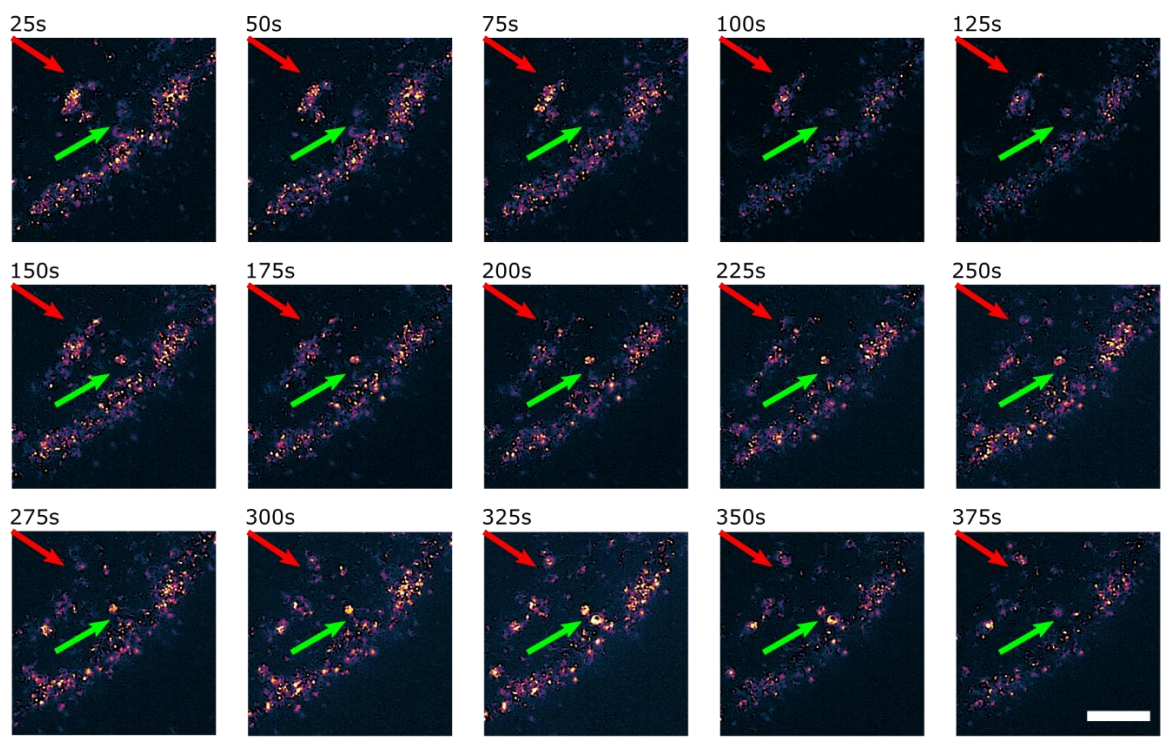

*Figure S9 Slow dynamics of wheat germ agglutinin-Abberior FLIP 565 stained U2OS cell.* 4th order SOFI close-up of areas marked in Figure 3 a at different time points with a sliding window of 50 s (1000 frames). Data is the same as in Figure 3. Scale bar 5μm.